\documentclass[prl,showpacs,twocolumn,preprintnumbers,amsmath,amssymb]{revtex4}
\usepackage{amsmath}
\usepackage{amsfonts}
\usepackage{amssymb}
\usepackage{color}

\usepackage{epsfig}
\usepackage{graphicx}
\usepackage{dcolumn}
\usepackage{bm}

\newcommand{\beq}{\begin{equation}}
\newcommand{\eeq}{\end{equation}}
\newcommand{\beqa}{\begin{eqnarray}}
\newcommand{\eeqa}{\end{eqnarray}}

\def\gapp{\lower.35em\hbox{$\stackrel{\textstyle>}{\sim}$}}
\def\lapp{\lower.35em\hbox{$\stackrel{\textstyle<}{\sim}$}}

\usepackage{ulem}
\begin{document}
\bibliographystyle{apsrev}
\title{Topological Fermi liquids from Coulomb interactions in the doped Honeycomb lattice}
\author{Eduardo V. Castro}
\affiliation{Instituto de Ciencia de Materiales de Madrid,\\
CSIC, Cantoblanco, E-28049 Madrid, Spain.}
\author{Adolfo G. Grushin}
\affiliation{Instituto de Ciencia de Materiales de Madrid,\\
CSIC, Cantoblanco, E-28049 Madrid, Spain.}
\author{Alberto Cortijo}
\affiliation{Departamento de F\'isica Te\'orica, Universidad Aut\'onoma de Madrid, E-28049, Madrid,
Spain}
\author{Fernando de Juan}
\affiliation{Department of Physics, Indiana University, Bloomington, IN 47405, USA}
\author{Bel\'en Valenzuela }
\affiliation{Instituto de Ciencia de Materiales de Madrid,\\
CSIC, Cantoblanco, E-28049 Madrid, Spain.}
\author{Mar\'{\i}a A. H. Vozmediano}
\affiliation{Instituto de Ciencia de Materiales de Madrid,\\
  CSIC, Cantoblanco, E-28049 Madrid, Spain.}
\date{\today}
\begin{abstract} 
We get an anomalous Hall metallic state in the Honeycomb lattice with nearest
neighbors only arising as a spontaneously broken symmetry state from a local
nearest neighbor Coulomb interaction $V$. 
The key ingredient is to enlarge the unit cell to host six atoms that permits
Kekul\'e distortions and supports self-consistent currents creating
non trivial magnetic configurations with total zero flux.
We find within a variational mean field approach
a metallic phase with broken time reversal symmetry ($\mathcal{T}$) 
very close in parameter space to a Pomeranchuk instability. 
Within the $\mathcal{T}$ broken region the 
predominant configuration is an anomalous Hall phase with non zero Hall conductivity, a realization of a topological Fermi liquid.
A $\mathcal{T}$ broken phase with zero Hall conductivity is stable in a small region
of the parameter space for lower values of $V$.

\end{abstract}
%

\maketitle

{\it Introduction.---}
The quantum Hall effect \cite{E08} and the anomalous Hall 
(AH) effect \cite{NSetal10} have given birth to 
a new paradigm in  condensed matter based on 
momentum space topology \cite{Vo03}. 
Very interesting developments have followed based on
the recent ideas of getting a 
Hall conductivity or Landau levels without external magnetic fields \cite{H88,GKG10}. 
These ideas have given rise to new areas 
of research and associated new materials as 
the topological insulators \cite{HK10} and the even more interesting
topological metals \cite{H04}. These systems allow the realization of beautiful fundamental ideas shared by different branches of physics like charge fractionalization or Majorana fermions 
\cite{HCM07,F10}.

The interplay of the underlying lattice and the electronic interactions
plays a very important role in the physics of these systems. In topological insulators
the spin-orbit coupling 
is the main ingredient to get non trivial topological phases. 
In the topological metals  time reversal symmetry  ($\mathcal{T}$) breaking without a magnetic field is the key ingredient, which can  be realized through  current (or bond) ordering: 
the electrons spontaneously form current loops, which interact among 
themselves in such a way that the state is self-consistently maintained. 
These phases were discussed in other contexts in
\cite{N91,L94,CLetal01,V06}.

Because of its special topology and long before 
the synthesis of graphene, the Honeycomb lattice has played a predominant role
in the modeling of topological states of matter.  
One of the earliest examples of a quantum Hall effect without a uniform externally applied magnetic  field is due to  Haldane \cite{H88} who obtained a 
$\mathcal{T}$
broken state in a tight  binding model in the Honeycomb 
lattice with complex values of the next to nearest neighbors
hopping parameters. Ever since, the search for realization
of spontaneous  (AH) effect has been very intense in the literature. The problem turns out to be very hard  
and the proposed models usually involve the inclusion of 
hopping or interactions beyond the nearest neighbors as in the original
Haldane model, or very elaborated lattice structures as the Kagom\'e or pyroclore 
\cite{SYFK09,RQetal08,PR10,WRetal10,Qetal10}.

\begin{figure}
\begin{centering}
\includegraphics[width=\columnwidth]{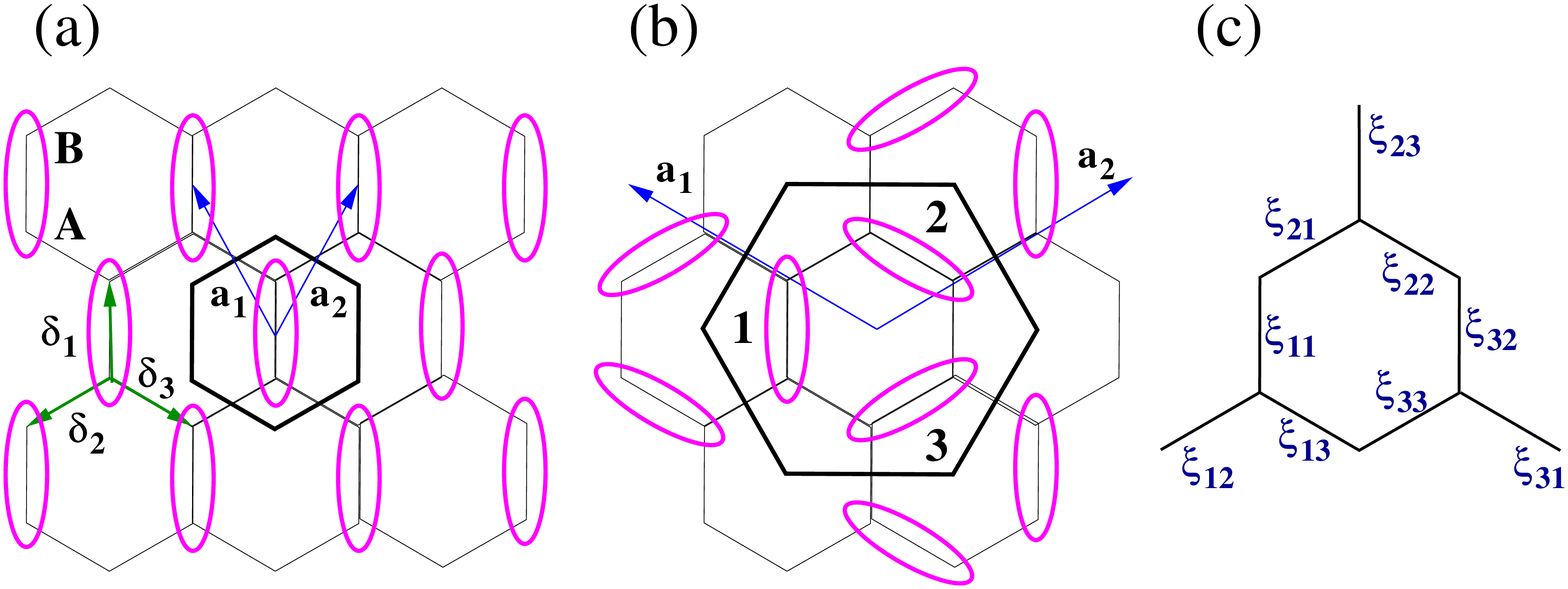}
\par\end{centering}
\caption{\label{fig:possibleDist}(color online). 
(a)~Two-atom unit cell and example of uniaxial distortion. (b)~Six atom
unit cell and Kekul\'e distortion, allowed in the enlarged unit cell.
(c)~ A pictorial representation of the nine complex order parameters  
considered in this work in the mean field decoupling of the Hamiltonian .}
\end{figure}

In this work we show that an AH phase
exists as a stable ground state in a simple nearest neighbor
tight binding model with nearest neighbor Coulomb interaction
in the Honeycomb lattice. The $\mathcal{T}$ 
broken state supporting this phase 
is made possible by enlarging the original two atom 
unit cell to a six atom unit cell (see Fig.~\ref{fig:possibleDist})
which allows for local non-zero current states 
and thus $\mathcal{T}$ broken phases. 
%
%
Our new phases are a realization of the topological Fermi
liquids described in  \cite{H04,SF08}. 

{\it The model.---}The extended Hubbard Hamiltonian that we consider for 
spinless fermions in the honeycomb lattice reads
\begin{equation}
H=-t\sum_{{{\bf r},{\bm \delta}}} a^\dagger_{{\bf r}} b_{{{\bf r} + {\bm \delta}}}
\;+\;
V\sum_{{\bf r}, {\bm \delta}}a^\dagger_{{\bf r }}a_{{\bf r }}
b^\dagger_{{\bf r } + {\bm \delta}}b_{{\bf r } + {\bm \delta}} 
\;+\;
h.c. \;, \label{ham-fs}
\end{equation}
where $t$ is the nearest neighbor hopping and $V$ the nearest 
neighbor Coulomb repulsion. We use standard notation where $a_{{\bf r }}$
($b_{{\bf r }}$) annihilates an electron at position ${\bf r}$ in sublattice
A (B). The two inequivalent sublattices A and B are depicted in 
Fig.~\ref{fig:possibleDist}(a), along with the basis vectors
$\mathbf{a}_{1}=\frac{\sqrt{3}a}{2}(-1,\sqrt{3})$ and 
$\mathbf{a}_{2}=\frac{\sqrt{3}a}{2}(1,\sqrt{3})$
for the case of a two atom unit cell. The vectors $\bm \delta$ refer to
the three vectors connecting nearest neighbor sites, as shown in 
Fig.~\ref{fig:possibleDist}(a).
To allow for $\mathcal{T}$ broken phases
as  mean field solutions we use an enlarged unit cell, containing six atoms, 
which also permits   Kekul\'e type distortion as shown in
Fig.~\ref{fig:possibleDist}(b). The basis
vectors of the enlarged cell in real space 
are $\mathbf{a}_{1}=\frac{3a}{2}(-\sqrt{3},1)$ and 
$\mathbf{a}_{2}=\frac{3a}{2}(\sqrt{3},1)$,
and the respective unit cell vectors in reciprocal space are
$\mathbf{b}_{1}=\frac{2\pi}{3\sqrt{3}a}(-1,\sqrt{3})$ and
$\mathbf{b}_{2}=\frac{2\pi}{3\sqrt{3}a}(1,\sqrt{3})$.
With this choice the unit cell in direct space is three times bigger
while in reciprocal space the Brillouin zone (BZ) becomes folded, i.e. it is three times 
smaller. This gives rise to a tight binding model whose wave function is a
six component spinor of the form 
\beq
\psi_{\mathbf{k}}^{\dagger}=[a_{1}^{\dagger}(\mathbf{k}),b_{1}^{\dagger}(\mathbf{k}),a_{2}^{\dagger}(\mathbf{k}),b_{2}^{\dagger}(\mathbf{k}),a_{3}^{\dagger}(\mathbf{k}),b_{3}^{\dagger}(\mathbf{k})].
\label{6spinor}
\eeq

Since we are interested in the electronic phases with broken $\mathcal{T}$ 
we do not consider for the time being charge ordered phases.  
Under these conditions the most general mean field Hamiltonian depends on 
nine complex parameters $\xi_{ij}$ which can be grouped in a $3\times 3$ 
matrix, and that can be shown to be $k-$independent.
The mean field equations can be written in terms of the mean field averages 
of the form 
$\langle b_{j}^{\dagger}(\mathbf{k})a_{i}(\mathbf{k})\rangle _{MF}$ as
\begin{equation}
\xi_{ij}=-\frac{2}{N}\sum_{\mathbf{k}}\gamma_{\mathbf{k}}^{ij}
\langle b_{j}^{\dagger}(\mathbf{k})a_{i}(\mathbf{k})\rangle _{MF},
\label{eq:SC}
\end{equation}
where $N$ is the number of unit cells, $\gamma_{\mathbf{k}}$ 
is a $3\times3$ matrix given by
\[
\gamma_{\mathbf{q}}=\left[\begin{array}{ccc}
1 & e^{-i\mathbf{a}_{2}\cdot\mathbf{k}} & 1\\      
1 & 1 & e^{i(\mathbf{a}_{1}+\mathbf{a}_{2})\cdot\mathbf{k}}\\
e^{-i\mathbf{a}_{1}\cdot\mathbf{k}} & 1 & 1\end{array}\right],\]
and the momentum sum runs over the folded BZ.
The nine complex order parameters $\xi_{ij}$ of our mean field decoupling 
represent the nine bonds in the enlarged unit cell, as pictorially
represented in Fig.~\ref{fig:possibleDist}(c).
We solve Eq.~(\ref{eq:SC}) self-consistently with the constrain
imposed by the Luttinger theorem \cite{L60}, 
which reads (ignoring logarithmic corrections in fermion number $N_{e}$),
\beq
n+3=\frac{N_{e}}{N} =  
\frac{1}{N}\sum_{\mathbf{k},l}n_F[\varepsilon_l(\mathbf{k}), \mu],
\label{eq:n}
\eeq
where $n$ is the electron density per unit cell relative to half filling (which
in our case corresponds to $n=0$),  
$\varepsilon_l (\mathbf{k})$ is the mean field dispersion for the $l$
band, and 
$n_F[\varepsilon_l(\mathbf{k}), \mu]$ is the Fermi distribution function.
From Eq.~\eqref{eq:n} we get the renormalized chemical potential 
$\mu$ self-consistently.

{\it The phase diagram and the AH phase.---}
\begin{figure}
\begin{center}
\includegraphics[width=\columnwidth]{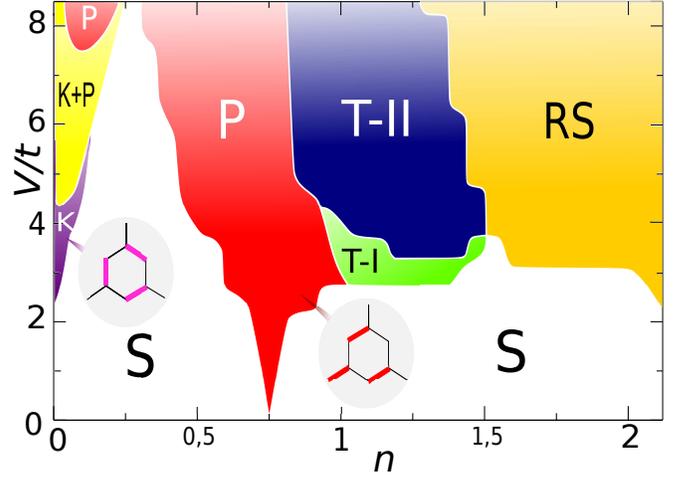}
\caption{(color online). Mean field phase diagram. Legend: (S) symmetric phase,
i.e. bare graphene with a uniform renormalization of the hopping;
(K) Kekul\'e distortion
with hopping renormalization as shown in the inset;
(P) Pomeranchuk distortion of the
Fermi surface and hopping renormalization as shown in the inset;
(K+P) coexistence of Kekule and Pomeranchuk distortions;
(T-I) and (T-II) anomalous Hall phases discussed at length in the text; 
(RS) broken symmetry state with real hopping parameters, the
distortion is neither Kekul\'e type nor Pomeranchuk (reduced symmetry).}
\label{fig:pd}
\end{center}
\end{figure}
The mean field phase diagram 
is shown in Fig.~\ref{fig:pd}
where we plot the different phases (defined in the caption)
as a function of 
the interaction strength $V$ in units of the hopping parameter $t$
and the electron density $n$.  The density varies 
from $n=0$ (half filling) to well above the VH filling which with our
convention occurs at $n_{VH}=0.75$ .
At each point in the phase diagram a mean field Hamiltonian can be
extracted which can be seen as a free Hamiltonian with new effective hopping parameters renormalized by the interaction. 
At low values of $V$ the symmetric phase (S) represents standard graphene with
a uniform renormalization of the hopping. Close to half filling for increasing values of $V$
slightly above $V=2t$ we recover the 
Kekul\'e phase (K) described in \cite{HCM07}
which evolves to a Pomeranchuk phase (P) through a finite coexistence 
region (K+P).
Our calculation shows  that a  standard Pomeranchuk
instability with an anisotropic renormalization 
of the hoppings as shown in the inset, 
is a very robust phase around the VH filling from zero to
high values of $V$. This
result was already obtained without the BZ folding in \cite{VV08,LCG09}. 
The preferred phase is a nematic one where the $C_6$ symmetry 
of the original lattice is broken to a $C_2$. 
The inset shows one of the three equivalent
configurations oriented along the crystal principal directions. 
The phase named reduced symmetry (RS) occurring at higher values of the electron 
density and the interaction is a
broken symmetry state with real hopping parameters. 
The distortion is neither Kekul\'e type nor Pomeranchuk.

\begin{figure}
\begin{center}
\includegraphics[width=0.8\columnwidth]{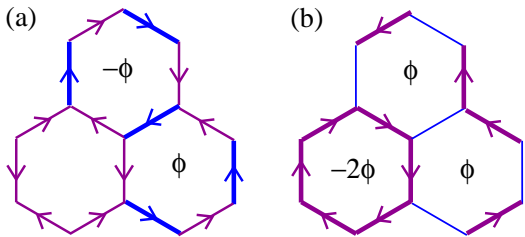}
\caption{\label{fig_fluxes}(color online). Pictorial 
representation of the order parameters corresponding to the $\mathcal{T}$ 
broken phases T-I (a) and T-II (b) discussed in the text. The thickness of the bonds
represent  the modulus of the hopping parameter 
and the direction of the arrows represents the sign of the phase when it has a complex value. A bond without an arrow means a real hopping.}
\end{center}
\end{figure}

The novel topological Fermi liquid phases appear near $n=1$.
There are two $\mathcal{T}$ broken phases labelled T-I and T-II 
in Fig.~\ref{fig:pd}
which are the most stable configurations
just above the VH filling for moderate values of $V$ beginning
at $V \approx 3t$. They are pictorially described in
Fig.~\ref{fig_fluxes}(a) and~\ref{fig_fluxes}(b), respectively, where
the nine complex order parameters of our mean field decoupling are shown. 
The direction of the arrows represents the sign of the phase of the given complex hopping, and the thickness of the line represents its modulus. The phases can also be understood as patterns of orbital currents. Current conservation at each of the six atoms in the unit cell plus the zero overall flux condition allow for only two independent $\mathcal{T}$ breaking phases, T-I and T-II in our notation, defined by their corresponding flux pattern in the unit cell. As it is clear from Fig. 3 both phases are only possible if the unit cell is enlarged. We note that  in addition of having the non-trivial fluxes described their structure includes a Kekul\'e distortion of the bonds. 

The discrete symmetries of the mean field Hamiltonian  help to classify the topological properties of a given phase \cite{SF08}. Following  \cite{MGV07} the discrete symmetry operations $\mathcal{T}$ and inversion symmetry ($\mathcal{I}$) amount to the transformation,
\beq
\mathcal{T}: \mathcal{H}_{\bf k}\to \mathcal{H}^*_{\bf -k} \qquad 
\mathcal{I}: \mathcal{H}_{\bf k}\to \mathcal{H}^{a_i \leftrightarrow b_{i+1}}_{\bf -k},
\label{eq:discrete}
\eeq
with $b_4 \equiv b_1$.
The phase T-I breaks $\mathcal{T}$ and $\mathcal{I}$, 
but preserves $\mathcal{TI}$. The T-II breaks
$\mathcal{T}$ but  $\mathcal{I}$ 
is preserved. 

In what follows we will show that the T-II phase that dominates the $\mathcal{T}$ broken part of the phase diagram is indeed an AH phase. 
As discussed in \cite{H04}
the (non-quantized) Hall conductivity of the topological metals as the ones encountered 
in this work
is a property of the Fermi surface. 
At a given point of the phase diagram 
it can be computed from the 
single particle Bloch states $|\Psi_l({\bf k})\rangle$
associated to the appropriate mean field Hamiltonian from the expression:
\begin{equation}
\label{AHC}
\sigma^{ab}(\mu)= \dfrac{e^2}{\hbar}\dfrac{1}{N \mathcal{V}}\sum_{k,l} 
\Omega^{ab}_{l}(\mathbf{k}) n_{F}[\varepsilon_l(\mathbf{k}),\mu],
\end{equation}
where  $\mathcal{V}$ is the volume of the unit cell, and
$ \Omega^{ab}_{l}(\mathbf{k}) $ is the Berry curvature defined from the
Berry connection:
${\cal A}_l^a({\bf k})=-i\left<\Psi_l({\bf k})\vert\nabla_{({\bf k})}^a\Psi_l({\bf k})\right>$,
$\Omega^{ab}_{l}(\mathbf{k})=\nabla_k^a{\cal A}_l^b({\bf k})-\nabla_k^b{\cal A}_l^a({\bf k}).$
The T-II phase  is of the type II in the classification given in \cite{SF08}:
it breaks $\mathcal{T}$ but preserves  $\mathcal{I}$ and the Hall conductivity 
is generically non zero.
The T-I phase, Fig.~\ref{fig_fluxes}(a), breaks $\mathcal{T}$ and $\mathcal{I}$
but preserves $\mathcal{TI}$ so it corresponds to a 
$\mathcal{T}$ broken phase of type I and has zero Hall conductivity. 
We have further confirmed this picture by  numerical computation of the Hall conductivity
Eq.~\eqref{AHC}.

\begin{figure}
\begin{center}
\includegraphics[height=4cm]{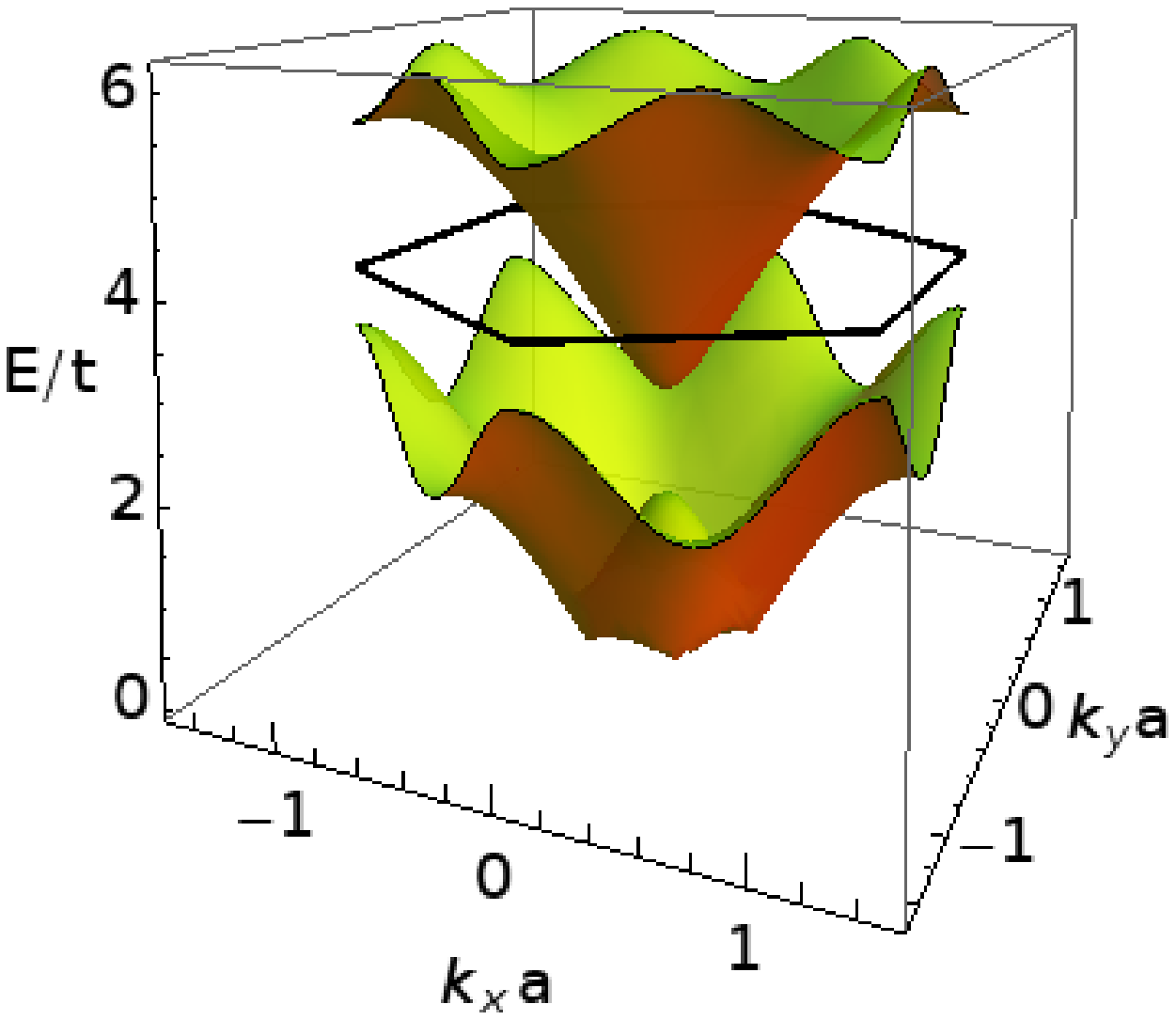}\;\;
\includegraphics[height=3cm]{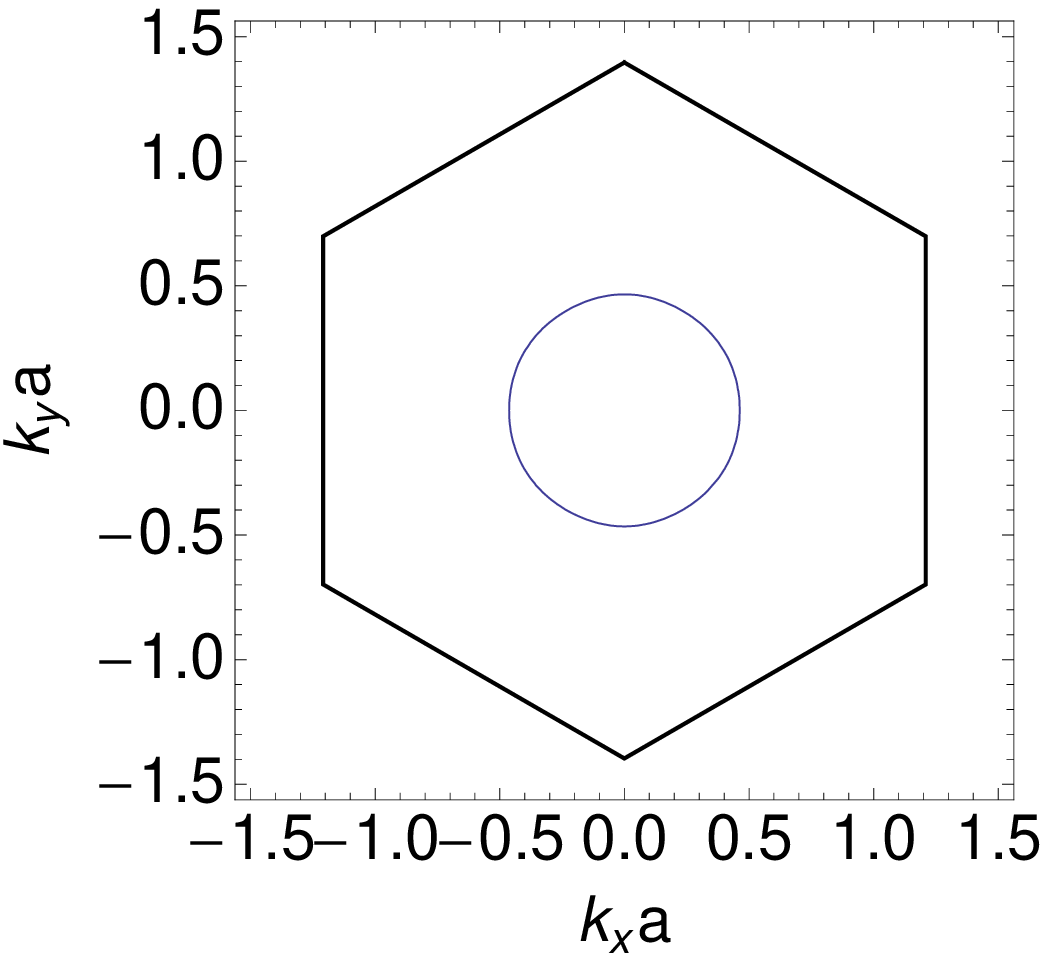}
\caption{\label{fig_FS_aH}(color online). 
(Left) Two lowest energy bands for the mean field Hamiltonian in the
AH phase obtained with $V=5t$, $n=1.13$. The hexagonal line marks the
position of the Fermi level. (Right) The Fermi surface.}
\end{center}
\end{figure}

A very neat analysis of the topological properties of the various 
metallic phases
in the phase diagram can be done by studying the low energy effective bands. 
We have plotted the mean field band structure of the T-II phase in  
Fig.~\ref{fig_FS_aH}
obtained for the parameter values $V=5t$, $n=1.13$. 
Focusing on the relevant bands around the Fermi level, it is easy to understand
the qualitative behavior of the non vanishing Hall conductivity in this phase.
The Fermi level crosses a massive Dirac structure around the $\Gamma$ point 
and so there is a non-zero contribution to the AH conductivity of this cone. 
The non-quantized contribution to the AH conductivity from the cone
is given by \cite{KM05b,CGV10}
\beq
\sigma_H=\frac{e^2}{2h}\frac{M(n,V)}{\vert\tilde\mu(n,V)\vert}
\eeq
where $M(n, V)$ is the gap at the $\Gamma$ point and 
$\tilde{\mu}(n, V)$ is the renormalized chemical potential relative
to the middle of the gap, both of which depend strongly on the parameters of the phase diagram.

To better understand the nature of the $\mathcal{T}$ broken phases we note that they arise in the region of the parameter space close to the density where there are four electrons per unit cell: $n=1$. This is a very special filling: Not only it is commensurate with the lattice, but it enhances the formation of current loops self-consistently maintained at each hexagon following the configurations shown in   Fig.~\ref{fig_fluxes}. We have seen that in the $\mathcal{T}$ broken part of the phase diagram, along  the line $n=1$ the system becomes an insulator. The band structure near the Fermi level is similar to the one shown in Fig. \ref{fig_FS_aH} left but the cones are further apart and the Fermi level lies in the gap.
Away from this line we have the situation described before. The majority of electrons will still form currents as these in Fig. \ref{fig_fluxes} and the excess (defect) electrons are responsible for the metallicity of the system. This picture allows also to understand the asymmetry of the phase diagram with respect to the line $n = 1$ (see Fig.~\ref{fig:pd}) as a result of the competition between the soft Fermi surface associated to the high density of states  near the VH filling and the orbital  ordering driven by high to moderate values of the interaction $V$ at the commensurate filling. Since the density of states (DOS)  is very steep around this density, moving to the left (for lower values of $n$) the DOS grows very fast towards the VH filling and the Pomeranchuk instability takes over. Going to densities $n>1$ the DOS decreases and the local currents responsible for the  $\mathcal{T}$ broken phases survive in a larger region.

{\it Discussion and future.---}
%
%
Part of the physics discussed in this work can be tested in actual graphene samples. The simple deformation of the Fermi surface pointing to a Pomeranchuk instability is a very robust phase that may prevail even if other instabilities not considered in this work are allowed. The AH phase can be more difficult to observe in graphene since it occurs at higher values of the interactions and is accompanied by a Kekul\'e distortion but it could potentially be tested in cold atom experiments with optical lattices \cite{SZ08,LGetal09}.

Other phases may compete with the ones described in this work 
when charge decoupling and spin degrees of freedom are included in the system.
It will be interesting to see how they compete with the AH phase obtained in this work.
Very appealing possibilities will open in the Pomeranchuk region
of the phase diagram in Fig.~\ref{fig:pd} when spin is included along 
the lines of \cite{WSetal07}. Spin effects  have also been explored recently in \cite{L11}. 
A detailed study of the physical properties of the various phases encountered as well as the detailed nature of the phase transitions will be examined in a forthcoming publication.

{\it Conclusions.---}
We have found a spontaneous symmetry breaking to 
an AH phase in a tight binding model
in the Honeycomb lattice with only nearest neighbor hopping parameters and
Coulomb interaction. The extra physics required to get such a phase
is provided by the folding of the BZ that allows for spontaneous non-zero currents with zero overall magnetic flux to form inside the unit cell generating 
$\mathcal{T}$ broken phases. 
The  $\mathcal{T}$ broken phase is predominantly an AH metal of the type II in the classification given in \cite{SF08} where the interaction $V$ gives rise to orbital current fluctuations  together with a  Kekul\'e distortion.

The findings of this work open a whole set of possibilities for new realization of exotic phases based on lattice models. Enlarging the unit cell is a very simple procedure that increases enormously the phase space of any given lattice. This is exemplified in the  model studied here where in addition to the AH phase we have found a very rich phase  diagram even when neglecting spin and charge instabilities. 

This research was supported by the
MEC (Spain) through grants FIS2008-00124, PIB2010BZ-00512. F.J. acknowledges support from NSF through Grant No. DMR-1005035.

\bibliography{TRB}

\newcommand{\npb}{Nucl. Phys. B}\newcommand{\adv}{Adv.
  Phys.}\newcommand{\epl}{Europhys. Lett.}
\begin{thebibliography}{29}
\expandafter\ifx\csname natexlab\endcsname\relax\def\natexlab#1{#1}\fi
\expandafter\ifx\csname bibnamefont\endcsname\relax
  \def\bibnamefont#1{#1}\fi
\expandafter\ifx\csname bibfnamefont\endcsname\relax
  \def\bibfnamefont#1{#1}\fi
\expandafter\ifx\csname citenamefont\endcsname\relax
  \def\citenamefont#1{#1}\fi
\expandafter\ifx\csname url\endcsname\relax
  \def\url#1{\texttt{#1}}\fi
\expandafter\ifx\csname urlprefix\endcsname\relax\def\urlprefix{URL }\fi
\providecommand{\bibinfo}[2]{#2}
\providecommand{\eprint}[2][]{\url{#2}}

\bibitem[{\citenamefont{Ezawa}(2008)}]{E08}
\bibinfo{author}{\bibfnamefont{Z.~F.} \bibnamefont{Ezawa}},
  \emph{\bibinfo{title}{Quantum Hall Effects - Field Theoretical Approach and
  Related Topics}} (\bibinfo{publisher}{World Scientific, Singapore},
  \bibinfo{year}{2008}).

\bibitem[{\citenamefont{Nagaosa et~al.}(2010)\citenamefont{Nagaosa, Sinova,
  Onoda, MacDonald, and Ong}}]{NSetal10}
\bibinfo{author}{\bibfnamefont{N.}~\bibnamefont{Nagaosa}},
  \bibinfo{author}{\bibfnamefont{J.}~\bibnamefont{Sinova}},
  \bibinfo{author}{\bibfnamefont{S.}~\bibnamefont{Onoda}},
  \bibinfo{author}{\bibfnamefont{A.~H.} \bibnamefont{MacDonald}},
  \bibnamefont{and} \bibinfo{author}{\bibfnamefont{N.~P.} \bibnamefont{Ong}},
  \bibinfo{journal}{Rev. Mod. Phys.} \textbf{\bibinfo{volume}{82}},
  \bibinfo{pages}{1539} (\bibinfo{year}{2010}).

\bibitem[{\citenamefont{Volovik}(2003)}]{Vo03}
\bibinfo{author}{\bibfnamefont{G.~E.} \bibnamefont{Volovik}},
  \emph{\bibinfo{title}{The universe in a helium droplet}}
  (\bibinfo{publisher}{Clarendon Press, Oxford}, \bibinfo{year}{2003}).

\bibitem[{\citenamefont{Haldane}(1988)}]{H88}
\bibinfo{author}{\bibfnamefont{F.~D.~M.} \bibnamefont{Haldane}},
  \bibinfo{journal}{Phys. Rev. Lett.} \textbf{\bibinfo{volume}{61}},
  \bibinfo{pages}{2015} (\bibinfo{year}{1988}).

\bibitem[{\citenamefont{Guinea et~al.}(2010)\citenamefont{Guinea, Katsnelson,
  and Geim}}]{GKG10}
\bibinfo{author}{\bibfnamefont{F.}~\bibnamefont{Guinea}},
  \bibinfo{author}{\bibfnamefont{M.~I.} \bibnamefont{Katsnelson}},
  \bibnamefont{and} \bibinfo{author}{\bibfnamefont{A.~G.} \bibnamefont{Geim}},
  \bibinfo{journal}{Nature Physics} \textbf{\bibinfo{volume}{6}},
  \bibinfo{pages}{30} (\bibinfo{year}{2010}).

\bibitem[{\citenamefont{Hasan and Kane}(2010)}]{HK10}
\bibinfo{author}{\bibfnamefont{M.~Z.} \bibnamefont{Hasan}} \bibnamefont{and}
  \bibinfo{author}{\bibfnamefont{C.~L.} \bibnamefont{Kane}},
  \bibinfo{journal}{Rev. Mod. Phys.} \textbf{\bibinfo{volume}{82}},
  \bibinfo{pages}{3045} (\bibinfo{year}{2010}).

\bibitem[{\citenamefont{Haldane}(2004)}]{H04}
\bibinfo{author}{\bibfnamefont{F.~D.~M.} \bibnamefont{Haldane}},
  \bibinfo{journal}{Phys. Rev. Lett.} \textbf{\bibinfo{volume}{93}},
  \bibinfo{pages}{206602} (\bibinfo{year}{2004}).

\bibitem[{\citenamefont{Hou et~al.}(2007)\citenamefont{Hou, Chamon, and
  Mudry}}]{HCM07}
\bibinfo{author}{\bibfnamefont{C.-Y.} \bibnamefont{Hou}},
  \bibinfo{author}{\bibfnamefont{C.}~\bibnamefont{Chamon}}, \bibnamefont{and}
  \bibinfo{author}{\bibfnamefont{C.}~\bibnamefont{Mudry}},
  \bibinfo{journal}{Phys. Rev. Lett.} \textbf{\bibinfo{volume}{98}},
  \bibinfo{pages}{186809} (\bibinfo{year}{2007}).

\bibitem[{\citenamefont{Franz}(2010)}]{F10}
\bibinfo{author}{\bibfnamefont{M.}~\bibnamefont{Franz}},
  \bibinfo{journal}{Physics} \textbf{\bibinfo{volume}{3}}, \bibinfo{pages}{24}
  (\bibinfo{year}{2010}).

\bibitem[{\citenamefont{Nersesyan}(1991)}]{N91}
\bibinfo{author}{\bibfnamefont{A.~A.} \bibnamefont{Nersesyan}},
  \bibinfo{journal}{Phys. Lett. A} \textbf{\bibinfo{volume}{153}},
  \bibinfo{pages}{49} (\bibinfo{year}{1991}).

\bibitem[{\citenamefont{Lieb}(1994)}]{L94}
\bibinfo{author}{\bibfnamefont{E.~H.} \bibnamefont{Lieb}},
  \bibinfo{journal}{Phys. Rev. Lett.} \textbf{\bibinfo{volume}{73}},
  \bibinfo{pages}{2158} (\bibinfo{year}{1994}).

\bibitem[{\citenamefont{Chakravarty et~al.}(2001)\citenamefont{Chakravarty,
  Laughlin, Morr, and Nayak}}]{CLetal01}
\bibinfo{author}{\bibfnamefont{S.}~\bibnamefont{Chakravarty}},
  \bibinfo{author}{\bibfnamefont{R.~B.} \bibnamefont{Laughlin}},
  \bibinfo{author}{\bibfnamefont{D.~K.} \bibnamefont{Morr}}, \bibnamefont{and}
  \bibinfo{author}{\bibfnamefont{C.}~\bibnamefont{Nayak}},
  \bibinfo{journal}{Phys. Rev. B} \textbf{\bibinfo{volume}{63}},
  \bibinfo{pages}{094503} (\bibinfo{year}{2001}).

\bibitem[{\citenamefont{Varma}(2006)}]{V06}
\bibinfo{author}{\bibfnamefont{C.~M.} \bibnamefont{Varma}},
  \bibinfo{journal}{Phys. Rev. B} \textbf{\bibinfo{volume}{73}},
  \bibinfo{pages}{155113} (\bibinfo{year}{2006}).

\bibitem[{\citenamefont{Sun et~al.}(2009)\citenamefont{Sun, Yao, Fradkin, and
  Kivelson}}]{SYFK09}
\bibinfo{author}{\bibfnamefont{K.}~\bibnamefont{Sun}},
  \bibinfo{author}{\bibfnamefont{H.}~\bibnamefont{Yao}},
  \bibinfo{author}{\bibfnamefont{E.}~\bibnamefont{Fradkin}}, \bibnamefont{and}
  \bibinfo{author}{\bibfnamefont{S.~A.} \bibnamefont{Kivelson}},
  \bibinfo{journal}{Phys. Rev. Lett.} \textbf{\bibinfo{volume}{103}},
  \bibinfo{pages}{046811} (\bibinfo{year}{2009}).

\bibitem[{\citenamefont{Raghu et~al.}(2008)\citenamefont{Raghu, Qi, Honerkamp,
  and Zhang}}]{RQetal08}
\bibinfo{author}{\bibfnamefont{S.}~\bibnamefont{Raghu}},
  \bibinfo{author}{\bibfnamefont{X.-L.} \bibnamefont{Qi}},
  \bibinfo{author}{\bibfnamefont{C.}~\bibnamefont{Honerkamp}},
  \bibnamefont{and} \bibinfo{author}{\bibfnamefont{S.-C.} \bibnamefont{Zhang}},
  \bibinfo{journal}{Phys. Rev. Lett.} \textbf{\bibinfo{volume}{100}},
  \bibinfo{pages}{156401} (\bibinfo{year}{2008}).

\bibitem[{\citenamefont{Pereg-Barnea and Refael}(2010)}]{PR10}
\bibinfo{author}{\bibfnamefont{T.}~\bibnamefont{Pereg-Barnea}}
  \bibnamefont{and} \bibinfo{author}{\bibfnamefont{G.}~\bibnamefont{Refael}},
  \bibinfo{journal}{arXiv:1011.5243}  (\bibinfo{year}{2010}).

\bibitem[{\citenamefont{Wen et~al.}(2010)\citenamefont{Wen, Ruegg, Wang, and
  Fiete}}]{WRetal10}
\bibinfo{author}{\bibfnamefont{J.}~\bibnamefont{Wen}},
  \bibinfo{author}{\bibfnamefont{A.}~\bibnamefont{Ruegg}},
  \bibinfo{author}{\bibfnamefont{C.-C.~J.} \bibnamefont{Wang}},
  \bibnamefont{and} \bibinfo{author}{\bibfnamefont{G.~A.} \bibnamefont{Fiete}},
  \bibinfo{journal}{Phys. Rev. B} \textbf{\bibinfo{volume}{82}},
  \bibinfo{pages}{075125} (\bibinfo{year}{2010}).

\bibitem[{\citenamefont{Qiao et~al.}(2010)}]{Qetal10}
\bibinfo{author}{\bibfnamefont{Z.}~\bibnamefont{Qiao}} \bibnamefont{et~al.},
  \bibinfo{journal}{Physical Review B} \textbf{\bibinfo{volume}{82}},
  \bibinfo{pages}{161414(R)} (\bibinfo{year}{2010}).

\bibitem[{\citenamefont{Sun and Fradkin}(2008)}]{SF08}
\bibinfo{author}{\bibfnamefont{K.}~\bibnamefont{Sun}} \bibnamefont{and}
  \bibinfo{author}{\bibfnamefont{E.}~\bibnamefont{Fradkin}},
  \bibinfo{journal}{Phys. Rev. B} \textbf{\bibinfo{volume}{78}},
  \bibinfo{pages}{245122} (\bibinfo{year}{2008}).

\bibitem[{\citenamefont{Luttinger}(1960)}]{L60}
\bibinfo{author}{\bibfnamefont{J.~M.} \bibnamefont{Luttinger}},
  \bibinfo{journal}{Phys. Rev.} \textbf{\bibinfo{volume}{119}},
  \bibinfo{pages}{1153} (\bibinfo{year}{1960}).

\bibitem[{\citenamefont{Valenzuela and Vozmediano}(2008)}]{VV08}
\bibinfo{author}{\bibfnamefont{B.}~\bibnamefont{Valenzuela}} \bibnamefont{and}
  \bibinfo{author}{\bibfnamefont{M.~A.~H.} \bibnamefont{Vozmediano}},
  \bibinfo{journal}{New J. Phys.} \textbf{\bibinfo{volume}{10}},
  \bibinfo{pages}{113009} (\bibinfo{year}{2008}).

\bibitem[{\citenamefont{Lamas et~al.}(2009)\citenamefont{Lamas, Cabra, and
  Grandi}}]{LCG09}
\bibinfo{author}{\bibfnamefont{C.}~\bibnamefont{Lamas}},
  \bibinfo{author}{\bibfnamefont{D.}~\bibnamefont{Cabra}}, \bibnamefont{and}
  \bibinfo{author}{\bibfnamefont{N.}~\bibnamefont{Grandi}},
  \bibinfo{journal}{Phys. Rev. B} \textbf{\bibinfo{volume}{80}},
  \bibinfo{pages}{075108} (\bibinfo{year}{2009}).

\bibitem[{\citenamefont{{J. Ma\~nes} et~al.}(2007)\citenamefont{{J. Ma\~nes},
  Guinea, and Vozmediano}}]{MGV07}
\bibinfo{author}{\bibnamefont{{J. Ma\~nes}}},
  \bibinfo{author}{\bibfnamefont{F.}~\bibnamefont{Guinea}}, \bibnamefont{and}
  \bibinfo{author}{\bibfnamefont{M.~A.~H.} \bibnamefont{Vozmediano}},
  \bibinfo{journal}{Phys. Rev. B} \textbf{\bibinfo{volume}{75}},
  \bibinfo{pages}{155424} (\bibinfo{year}{2007}).

\bibitem[{\citenamefont{Kane and Mele}(2005)}]{KM05b}
\bibinfo{author}{\bibfnamefont{C.}~\bibnamefont{Kane}} \bibnamefont{and}
  \bibinfo{author}{\bibfnamefont{E.}~\bibnamefont{Mele}},
  \bibinfo{journal}{Phys, Rev. Lett.} \textbf{\bibinfo{volume}{95}},
  \bibinfo{pages}{146802} (\bibinfo{year}{2005}).

\bibitem[{\citenamefont{Cortijo et~al.}(2010)\citenamefont{Cortijo, Grushin,
  and Vozmediano}}]{CGV10}
\bibinfo{author}{\bibfnamefont{A.}~\bibnamefont{Cortijo}},
  \bibinfo{author}{\bibfnamefont{A.~G.} \bibnamefont{Grushin}},
  \bibnamefont{and} \bibinfo{author}{\bibfnamefont{M.~A.~H.}
  \bibnamefont{Vozmediano}}, \bibinfo{journal}{Phys. Rev. B}
  \textbf{\bibinfo{volume}{82}}, \bibinfo{pages}{195438}
  (\bibinfo{year}{2010}).

\bibitem[{\citenamefont{Shao et~al.}(2008)\citenamefont{Shao, Zhu, Sheng, Xing,
  and Wang}}]{SZ08}
\bibinfo{author}{\bibfnamefont{L.~B.} \bibnamefont{Shao}},
  \bibinfo{author}{\bibfnamefont{S.-L.} \bibnamefont{Zhu}},
  \bibinfo{author}{\bibfnamefont{L.}~\bibnamefont{Sheng}},
  \bibinfo{author}{\bibfnamefont{D.~Y.} \bibnamefont{Xing}}, \bibnamefont{and}
  \bibinfo{author}{\bibfnamefont{Z.~D.} \bibnamefont{Wang}},
  \bibinfo{journal}{Phys. Rev. Lett.} \textbf{\bibinfo{volume}{101}},
  \bibinfo{pages}{246810} (\bibinfo{year}{2008}).

\bibitem[{\citenamefont{Lee et~al.}(2009)\citenamefont{Lee, Gremaud, R-Han,
  Englert, and Miniatura}}]{LGetal09}
\bibinfo{author}{\bibfnamefont{K.~L.} \bibnamefont{Lee}},
  \bibinfo{author}{\bibfnamefont{B.}~\bibnamefont{Gremaud}},
  \bibinfo{author}{\bibnamefont{R-Han}}, \bibinfo{author}{\bibfnamefont{B.-G.}
  \bibnamefont{Englert}}, \bibnamefont{and}
  \bibinfo{author}{\bibfnamefont{C.}~\bibnamefont{Miniatura}},
  \bibinfo{journal}{Phys. Rev. A} \textbf{\bibinfo{volume}{80}},
  \bibinfo{pages}{043411} (\bibinfo{year}{2009}).

\bibitem[{\citenamefont{Wu et~al.}(2007)\citenamefont{Wu, Sun, Fradkin, and
  Zhang}}]{WSetal07}
\bibinfo{author}{\bibfnamefont{C.}~\bibnamefont{Wu}},
  \bibinfo{author}{\bibfnamefont{K.}~\bibnamefont{Sun}},
  \bibinfo{author}{\bibfnamefont{E.}~\bibnamefont{Fradkin}}, \bibnamefont{and}
  \bibinfo{author}{\bibfnamefont{S.~C.} \bibnamefont{Zhang}},
  \bibinfo{journal}{Phys. Rev. B} \textbf{\bibinfo{volume}{75}},
  \bibinfo{pages}{115103} (\bibinfo{year}{2007}).

\bibitem[{\citenamefont{Li}(2011)}]{L11}
\bibinfo{author}{\bibfnamefont{T.}~\bibnamefont{Li}},
  \bibinfo{journal}{arXiv:1103.2420}  (\bibinfo{year}{2011}).

\end{thebibliography}
\end{document}